\documentclass[aps,prl,reprint,superscriptaddress,amsfonts, amsmath,
amssymb,showpacs]{revtex4-1}

\usepackage{bbm}

\usepackage{layout}

\usepackage{graphicx}
\newcommand{\dd}{{\rm d}}

\newcommand{\bsa}{{\mathbf a}}
\newcommand{\bsf}{{\mathbf f}}
\newcommand{\bsp}{{\mathbf p}}
\newcommand{\bsr}{{\mathbf r}}
\newcommand{\bsv}{{\mathbf v}}

\newcommand{\bsA}{{\mathbf A}}
\newcommand{\bsD}{{\mathbf D}}

\newcommand{\bsZ}{{\mathbf Z}}

\newcommand{\bGa}{{\boldsymbol \Gamma}}

\newcommand{\cF}{{\mathcal F}}

\newcommand{\bsone}{{\mathbf 1}}

\newcommand{\grad}[1]{{\boldsymbol \nabla}_{#1}}
\renewcommand{\div}[1]{\boldsymbol \nabla_{#1} \cdot}

\newcommand{\fund}[2]{\frac{\delta{#1}}{\delta {#2}}}

\DeclareMathOperator{\erf}{erf}

\begin{document}

\title{General dynamical density functional theory for classical fluids}

\author{Benjamin D.\ Goddard}
\affiliation{Department of Chemical Engineering,
Imperial College London, London, SW7 2AZ, United Kingdom}

\author{Andreas Nold}
\affiliation{Department of Chemical Engineering,
Imperial College London, London, SW7 2AZ, United Kingdom}
\affiliation{Center of Smart Interfaces, TU Darmstadt,
 Petersenstr.\ 32, 64287 Darmstadt, Germany}

\author{Nikos Savva}
\affiliation{Department of Chemical Engineering,
Imperial College London, London, SW7 2AZ, United Kingdom}

\author{Grigorios A.\ Pavliotis}
\affiliation{Department of Mathematics, Imperial
College London, London, SW7 2AZ, United Kingdom}
\affiliation{Institut f\"ur Mathematik, Freie Universit\"at Berlin,
Arnimallee 6, 14195 Berlin, Germany}

\author{Serafim Kalliadasis}
\affiliation{Department of Chemical Engineering,
Imperial College London, London, SW7 2AZ, United Kingdom}

\date{\today}

\begin{abstract}
We study the dynamics of a colloidal fluid including
inertia and hydrodynamic interactions, two effects which strongly
influence the non-equilibrium properties of the system. We derive a
general dynamical density functional theory (DDFT) which shows very
good agreement with full Langevin dynamics.
In suitable limits, we
recover existing DDFTs and a Navier-Stokes-like equation with
additional non-local terms.
\end{abstract}


\maketitle

Since the observation of the Brownian motion of pollen
particles in water in the 19th Century~\cite{Brown28}, the study of
classical fluids has been fundamental not only to the development of
statistical mechanics~\cite{Einstein05,*Langevin08,*Smoluchowski15},
but also to many other fields in physics, chemistry and engineering,
e.g. the evolution of microscopy over the last
century~\cite{Perrin09,*Franosch11,*Huang11}, recent advances in
biophysical research~\cite{Matijevic08} and the rapidly-growing field
of microfluidics~\cite{Evans99,*Stone2001,*Caruso2004}.

Colloidal systems, in particular, are versatile model ones for both
theoretical and experimental scrutiny. Many of the forces governing
their structure and behaviour govern also those of matter, whilst
the sufficiently large physical size of colloidal particles makes
them accessible experimentally. However, the large number of
particles in real-world systems translates to high-dimensional
mathematical models, which quickly become computationally
intractable.

Non-equilibrium statistical mechanics
approaches~\cite{Kreuzer81,Lutsko10}, such as the Boltzmann equation, allow the
dynamics of systems of arbitrarily large numbers of particles to be studied.
An important example is dynamical density
functional theory (DDFT)~\cite{Lutsko10} for the evolution of the one-body mass
distribution.
However, existing DDFTs neglect either
the momentum of the colloidal
particles~\cite{RexLowen09}, or the hydrodynamic interactions (HI)
mediated through the bath~\cite{Archer09}, or both, as in the
pioneering work in~\cite{MarconiTarazona99}. Yet, inertial effects
are negligible only in the high-friction limit~\cite{GPK11}, whilst
HI are long range~\cite{Dhont96}; it is therefore unclear that
existing DDFTs are sufficiently accurate to model general colloidal
systems. Here we outline a DDFT formalism which carefully and systematically
accounts for inertia and HI, an important step
towards accurate and predictive modelling of physically-relevant
systems.  It is validated with stochastic simulations,
and existing
DDFTs~\cite{MarconiTarazona99,RexLowen09,Archer09} are shown to be
special cases.

We are interested in systems
with a
large number $N$
identical, spherically symmetric colloidal particles of mass $m$
suspended in a bath of many more, much smaller and much lighter
particles.  Typically, colloidal particles are of size
1nm--1$\mu$m, occupying the same volume as approximately
$10^7$--$10^{10}$ water molecules.  As such, treating the bath
particles exactly is computationally prohibitive.  However, a
typical timescale for a colloidal particle to diffuse a distance
equal to its diameter is $1$s, whilst the typical time between
collisions of water molecules is $\tau_b \approx 10^{-15}$s~
\cite{PaddingLouis06}. Hence, for timescales
significantly larger than $\tau_b$, we may
introduce a coarse-grained model and consider only the colloidal particles, 
treating the bath in a
stochastic manner.

This approximation leads to the Langevin~\cite{Langevin08}
equations for the
$3N$-dimensional colloidal position and momentum vectors
$\bsr=(\bsr_1,\dots,\bsr_N)$ and
$\bsp=(\bsp_1,\dots,\bsp_N)$ with $\bsr_i$ and $\bsp_i$ the position and
momentum of the $i$th particle:
\begin{equation}
\frac{\dd \bsr}{\dd t} = \frac{\bsp}{m}, \quad
\frac{\dd \bsp}{\dd t} = -\grad{\bsr}V(\bsr,t) - \bGa (\bsr) \bsp +
\bsA(\bsr) \bsf(t). \label{Newton}
\end{equation}
Here, $V$ is the potential, generally a sum of an external
potential, such as gravity, and inter-particle potentials, such as
electrostatic effects. The motion of the colloidal particles causes
flows in the bath, which in turn cause forces on the colloidal
particles, referred to already as HI. The momenta and forces are
related by the $3N \times 3N$ positive-definite friction tensor
$\bGa$. See Supplemental Material for demonstrations of these
effects on sedimenting hard spheres.
Finally, collisions of bath particles with colloidal particles are
described by stochastic forces $\bsf$, given by Gaussian white
noise, the strength of which is determined by a generalized
fluctuation-dissipation theorem~\cite{ErmakMcCammon78}, $\bsA(\bsr)=(m
k_BT \bGa(\bsr))^{1/2}$,
with $T$ the temperature and $k_B$
Boltzmann's constant. We assume that $T$ is constant in space, i.e.\
that the solvent bath is also a heat bath on colloidal timescales.

When $N$ is large, interest lies not in particular realizations of
\eqref{Newton}, or experiments, but in averages over
a large number of them.
Averaging \eqref{Newton} over the initial
particle distribution and the noise leads to the Kramers (Fokker-Planck) equation, a
$6N$-dimensional deterministic PDE for the evolution
of the distribution function $f^{(N)}(\bsr,\bsp,t)$, the
probability of finding the particles with positions $\bsr$ and
momenta $\bsp$ at time $t$:
\begin{align}
& \big[\partial_t + \frac{1}{m} \bsp \cdot \grad{\bsr} -
 \grad{\bsr} V(\bsr,t) \cdot \grad{\bsp} \big] f^{(N)}(\bsr,\bsp,t)
 \label{Kramers}  \\
&\qquad - \div{\bsp} [ \bGa(\bsr)(\bsp + mk_BT
\grad{\bsp}) f^{(N)}(\bsr,\bsp,t) ] =0\notag
\end{align}

The main issue with solving \eqref{Newton} or \eqref{Kramers}, as with any
molecular approach, is that of
computational intensity for large systems. For \eqref{Kramers},
taking $M$ discretization points for each dimension
would require $M^{6N}$ points.  Hence, the only way to
solve \eqref{Kramers} for many particles is via Monte Carlo methods,
i.e. by solving \eqref{Newton}. However, for non-trivial HI this
requires $\mathcal{O}(N^3)$ operations at each timestep,
prohibiting calculations for many-particle systems. (Additionally, the
characteristic scale of the spatial structures is often too
large to be accurately treated.)

In contrast, it is known rigorously~\cite{ChanFinken05} that the $N$-body 
distribution function
$f^{(N)}$ is a functional of the one-body position distribution
$\rho(\bsr_1,t)= N\int \dd \bsp \dd \bsr' f^{(N)}(\bsr,\bsp,t)$,
where $\dd \bsr'$ denotes integration over all positions except
$\bsr_1$. Hence, for any number of particles, the
system is, in principle, completely described by a function
of only a single three-dimensional position variable (cf.\
TDDFT in quantum mechanics~\cite{Marques06}).

This motivates the derivation of a DDFT, 
a closed evolution equation for $\rho$.
We consider the moments of \eqref{Kramers} with respect to
momentum and obtain an infinite hierarchy of equations, which must be
truncated. This is
analogous to deriving the Euler or Navier-Stokes equation from the
Boltzmann equation~\cite{ResiboisDeLeener77}. We truncate the
hierarchy at the second equation; the next level treats the local temperature,
which here is constant due to the heat bath.
However, if required, this method can be systematically extended to higher
levels of the hierarchy.

We obtain a continuity equation for the density
\begin{equation}
\partial_t \rho(\bsr_1,t) + \div{\bsr_1}\big( \rho(\bsr_1,t)
\bsv(\bsr_1,t) \big) =0 \label{continuity}
\end{equation}
and an evolution equation for the local velocity
$\bsv(\bsr,t)=m^{-1}\int
\dd \bsp \dd \bsr' \bsp f^{(N)}(\bsr,\bsp,t)$:
\begin{align}
& D_t \bsv(\bsr_1,t) +
\frac{1}{\rho(\bsr_1,t)} \div{\bsr_1} \int \dd \bsp_1 \frac{\bsp_1 \otimes
\bsp_1}{m^2} f_{\rm neq}^{(1)}(\bsr_1,\bsp_1,t) \notag \\
 =  &- \frac{1}{m} \grad{\bsr_1} \fund{\cF[\rho]}{\rho} - \gamma
\bsv(\bsr_1,t) \label{DDFT}\\
& -  \gamma \int \dd \bsr_2  \rho(\bsr_2,t) g(\bsr_1,\bsr_2,[\rho])
   \sum_{j=1}^2 \bsZ_j(\bsr_1,\bsr_2) \bsv(\bsr_j,t)  \notag  
\end{align}
Here $D_t = \partial_t + \bsv(\bsr_1,t)\cdot \grad{\bsr_1}$ is the material
derivative and $\cF$ is the (equilibrium) Helmholtz free energy functional; see
later. 
For ease of
exposition, we restrict to two-body HI:
$\bGa(\bsr)= \gamma[
\bsone + \tilde \bGa (\bsr)]$, with the HI tensor $\tilde \bGa$
decomposed into $3\times 3$ blocks
$
    \tilde \bGa_{ij}(\bsr) = \delta_{ij} \sum_{\ell \neq i}
    \bsZ_1(\bsr_i,\bsr_\ell) + (1-\delta_{ij}) \bsZ_2(\bsr_i,\bsr_j)
$
\cite{GPK11}.  Here
$\bsone$ is the $3N \times 3N$ identity matrix and $\gamma$ is the friction
felt by a single, isolated particle.
Physically, $\tilde \bGa_{ij}$ describes the force on particle $i$ due to
the momentum of particle $j$. 
This two-body formulation is generally more accurate than that for
the diffusion tensor (as in \cite{RexLowen09}), which can lead to incorrect
physics \cite{KnudsenWerthWolf08}.
We have decomposed the one-body
distribution $f^{(1)}(\bsr_1,\bsp_1,t) =
N\int \dd \bsp' \dd \bsr' f^{(N)}(\bsr,\bsp,t) = f^{(1)}_{\rm
le}(\bsr_1,\bsp_1,t) +
f^{(1)}_{\rm neq}(\bsr_1,\bsp_1,t), $ where $f_{\rm le}^{(1)}$ is
the local-equilibrium part, the momentum dependence of which is
given by a local Maxwellian with mass $\rho(\bsr,t)$, mean
$m\rho(\bsr,t)\bsv(\bsr,t)$ and variance $mk_BT\rho(\bsr,t)$. 
The corresponding three quantities are zero for the non-equilibrium part
$f_{\rm neq}^{(1)}$.
We have also written the two-body reduced
distribution as
$f^{(2)}(\bsr_1,\bsr_2,\bsp_1,\bsp_2,t) =
f^{(1)}(\bsr_1,\bsp_1,t)f^{(1)}(\bsr_1,
\bsp_1,t)g(\bsr_1,\bsr_2,[\rho])$~\cite{Kreuzer81}.

The non-local terms in \eqref{DDFT}, absent from previous DDFTs, model important
physical effects.
That involving $\bsZ_1$ combines with $\gamma \bsv$ to give an
effective, density-dependent friction coefficient.  The
$\bsZ_2$ term non-locally couples the velocities.  Surprisingly, this
does not require explicit momentum correlations in $g$.
Neglecting these terms and setting $f^{(1)}_{\rm neq}=0$ recovers a
previous DDFT~\cite{Archer09}.
Setting $\gamma=0$
gives a DDFT for atomic and molecular fluids, although 
the closures below are harder to justify.

The non-trivial challenge here is
to close the momentum equation (see Supplemental Material) as a
functional of $\rho$.  We briefly describe three steps:

At equilibrium there exists an
exact functional identity~\cite{Evans79}
   $
   N\int \dd \bsr' \grad{\bsr_1} V(\bsr)
      \rho^{(N)}(\bsr) = \big(\grad{\bsr_1}\fund{\cF[\rho]}{\rho}
    - k_BT\grad{\bsr_1} \big) \rho(\bsr_1),
    $
where
$\cF[\rho] =k_BT\int \dd \bsr_1 \rho(\bsr_1) \big[ \ln
\big(\Lambda^3 \rho(\bsr_1) \big)-1 \big] + \cF_{\rm exc}[\rho] +
\int \dd \bsr_1 \rho(\bsr_1) V_{1}(\bsr_1) $ with $\Lambda$ the (irrelevant) de
Broglie wavelength. In
general, $\cF_{\rm exc}$ (the excess over ideal gas term) is unknown
but has been well-studied at equilibrium and good approximations
exist, e.g.\ fundamental measure theory~\cite{Rosenfeld89,Lutsko10}
(accurate for hard spheres) and mean field theory~\cite{Lutsko10},
(exact for soft interactions at high densities). 
We thus assume that the same identity holds
out of equilibrium, in particular giving the correct equilibrium
behaviour.

Since HI vanish at equilibrium there exists no analogous identity.  
Instead, we assume the form for $f^{(2)}$ given above for a
\emph{known} functional $g$. To
go beyond this two-body approach it is necessary to obtain
higher-order reduced distributions as functionals of $\rho$.

The term in \eqref{DDFT} containing $f^{(1)}_{\rm neq}$ is analogous to the
kinetic pressure tensor~\cite{Kreuzer81},
and there is no reason to expect it to be a simple functional of $\rho$
and $\bsv$. However, if it
may be neglected (e.g.\ via a maximum entropy approach
\cite{HughesBurghardt12}) or approximated as
a functional of $\rho$ and $\bsv$ (e.g.\ via a
Chapman-Enskog expansion), \eqref{continuity} and \eqref{DDFT} give a
DDFT. 
Alternatively, extending the above hierarchy removes the need for this
approximation, at the expense of requiring one for a higher-order moment of
$f^{(1)}$.

Since these approximations are unconstrained, it is crucial to test
them numerically. As far as we know, these are the first
such verifications of a phase space DDFT.
We now describe three such tests for hard
spheres of diameter $\sigma$.  We non-dimensionalize the
equations with the units of length, mass and energy being $\sigma$, $m$ and
$k_BT$ respectively.  We set $f_{\rm neq}^{(1)}=0$, use the hard-sphere FMT
functional~\cite{Rosenfeld89}, and  choose $g$ to be the
simplest possible (volume-exclusion) pair correlation function
$g(\bsr_1,\bsr_2,[\rho])=1$ for $|\bsr_1-\bsr_2|>1$ and zero
otherwise. Whilst not
entirely consistent with the FMT approximation, this s sufficiently accurate for
our purposes.
For HI we choose the
Rotne-Prager approximation~\cite{RotnePrager69,RexLowen09} in the overdamped
limit and its inverse for $\bGa$ in \eqref{Newton}.  We use an 11-term
two-body expansion~\cite{JeffreyOnishi84} for $\bGa$ in \eqref{DDFT}, leading to small quantitative differences between \eqref{Newton} and \eqref{DDFT}.
See Supplemental Material.

We take external potentials which depend spatially only on
$|\bsr_1|$, and assume that the same holds for $\rho$ and
$\bsv$, giving a 1D DDFT
problem.  We use spectral methods~\cite{Boyd01}, appropriately
extended to integral operators and a fifth order implicit
Runge-Kutta method with step size control~\cite{HairerWanner96}. 
The infinite physical domain is
mirrored and algebraically mapped onto $[-1,1]$ with 200 Chebyshev collocation
points, avoiding the singularity at the origin.
To capture the exponential decay of $\rho$, \eqref{continuity} and
\eqref{DDFT} are reformulated for $\log \rho + V_1$. The initial condition is
obtained from equilibrium DFT~\cite{Evans79}. We solve the
stochastic equations via an
Euler-Maruyama scheme with $10^5$ time steps,
averaged over 5000 runs, with initial conditions chosen via
slice sampling the (unnormalized) equilibrium $N$-body distribution.
The hard sphere potential is approximated via a slightly softened,
differentiable one~\cite{RexLowen09}.

\begin{figure}%[htb]
% l b r t
\includegraphics[width=0.45\textwidth,
height=0.25\textheight]{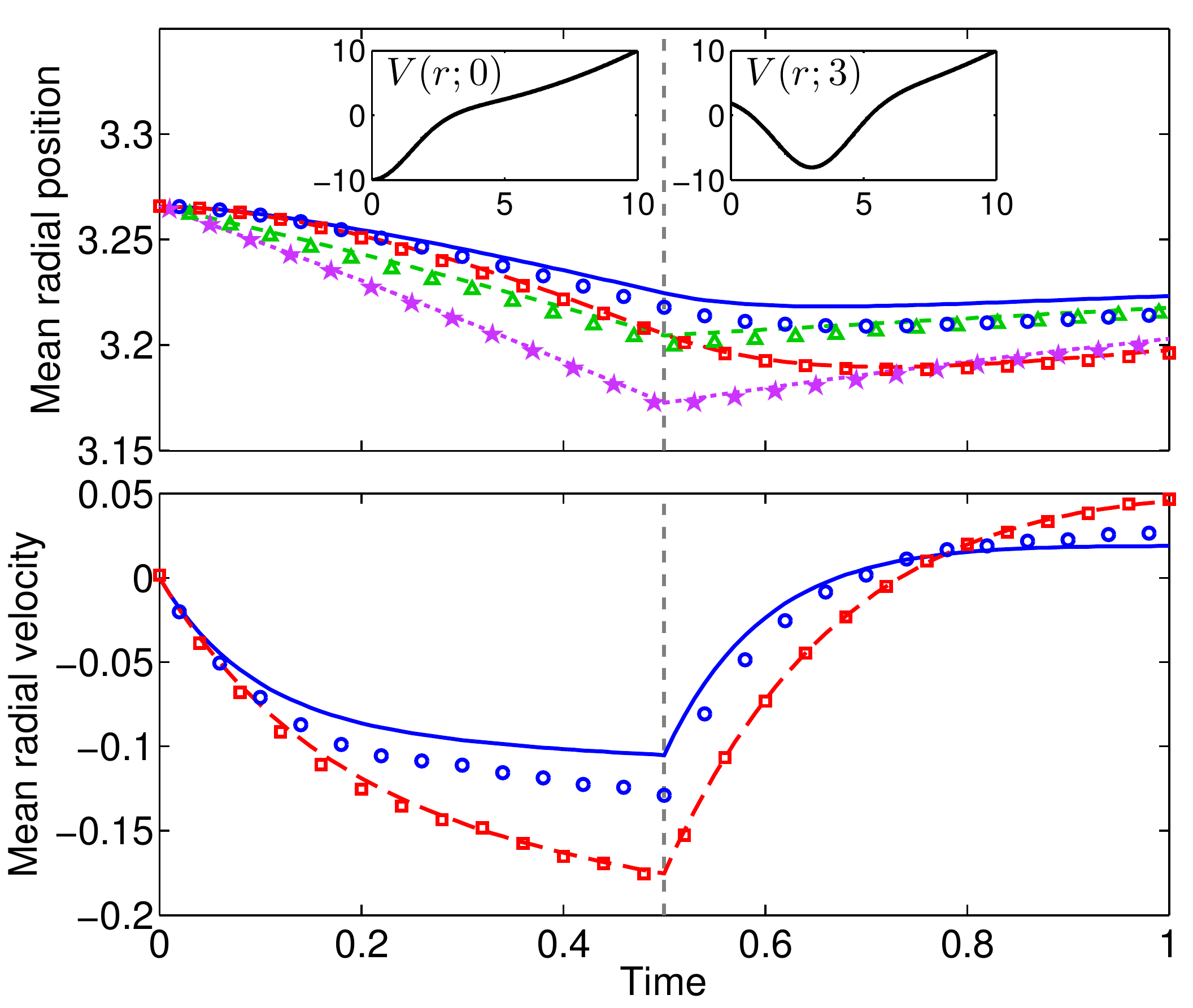} \caption{
\label{Fig:means} Mean radial positions and velocities from DDFT
\emph{(lines)} and stochastic equations \emph{(symbols)}. Full phase
space with \emph{(blue, solid, circles}) and without \emph{(red, long dashes,
squares)} HI from DDFT
\eqref{continuity} and \eqref{DDFT} and stochastics \eqref{Newton}.
Overdamped limit DDFT~\cite{RexLowen09} and
stochastics~\cite{ErmakMcCammon78} with \emph{(green, short dashes, triangles)}
and without
\emph{(purple, dots, stars)} HI.
}
\end{figure}

Fig.~\ref{Fig:means} shows the mean radial position and
velocity of 50 particles, with $\gamma=6$, starting at equilibrium
in a radially-symmetric external potential $V_1(r;3)$ with
$    V_1(r;r_0)=0.1(1-h)r^2 + 3h - 10 \exp[ -(r-r_0)^2/4 ]$,
where
$h(r)=[\erf((r+r_0)/2)-\erf((r-r_0)/2)]/2$ is a smooth cutoff. The
potential is
instantaneously
switched to $V_1(r;0)$ at time 0, and back to
$V_1(r;3)$ at time 0.5. The choice of 50 particles is large enough to overcome
the differences between the canonical ensemble stochastic and grand
canonical ensemble DDFT models, but also allows ease of access to
stochastic simulations. We show four pairs of computations, each
containing the solutions of a DDFT (lines) and the corresponding
stochastic equation (symbols). The first pair (blue, solid) includes
both
inertia and HI and compares our DDFT
\eqref{continuity} and \eqref{DDFT} to the
Euler-Maruyama~\cite{KloedenPlaten92} solution of \eqref{Newton} (circles). The
second pair (red, long dashes, squares) are the same simulations, but when HI
are neglected by setting $\bGa=\gamma \bsone$; see~\cite{Archer09}.
The
agreement between the DDFTs and stochastic simulations is very
good. The HI effects are quite
striking; they increase the effective friction and
damp the dynamics.

The remaining two pairs of simulations in Fig.~\ref{Fig:means} are restricted to
position
space via the high-friction approximation.
The DDFTs both with~\cite{RexLowen09} (green, short dashes) and
without~\cite{MarconiTarazona99} (purple, dots) HI, are compared to the
Ermak-McCammon~\cite{ErmakMcCammon78} solution of the corresponding
stochastic equations (triangles, stars respectively).  Whilst the agreement
between DDFT and
stochastic simulations is again very good, neglecting inertia leads to
qualitatively different behaviour
of the system, resulting in a kink in the mean position,
compared to smooth curves with a delay before the
mean velocity changes sign. Again, HI are seen to significantly
damp the dynamics.

\begin{figure}%[htb]
\includegraphics[width=0.43\textwidth]{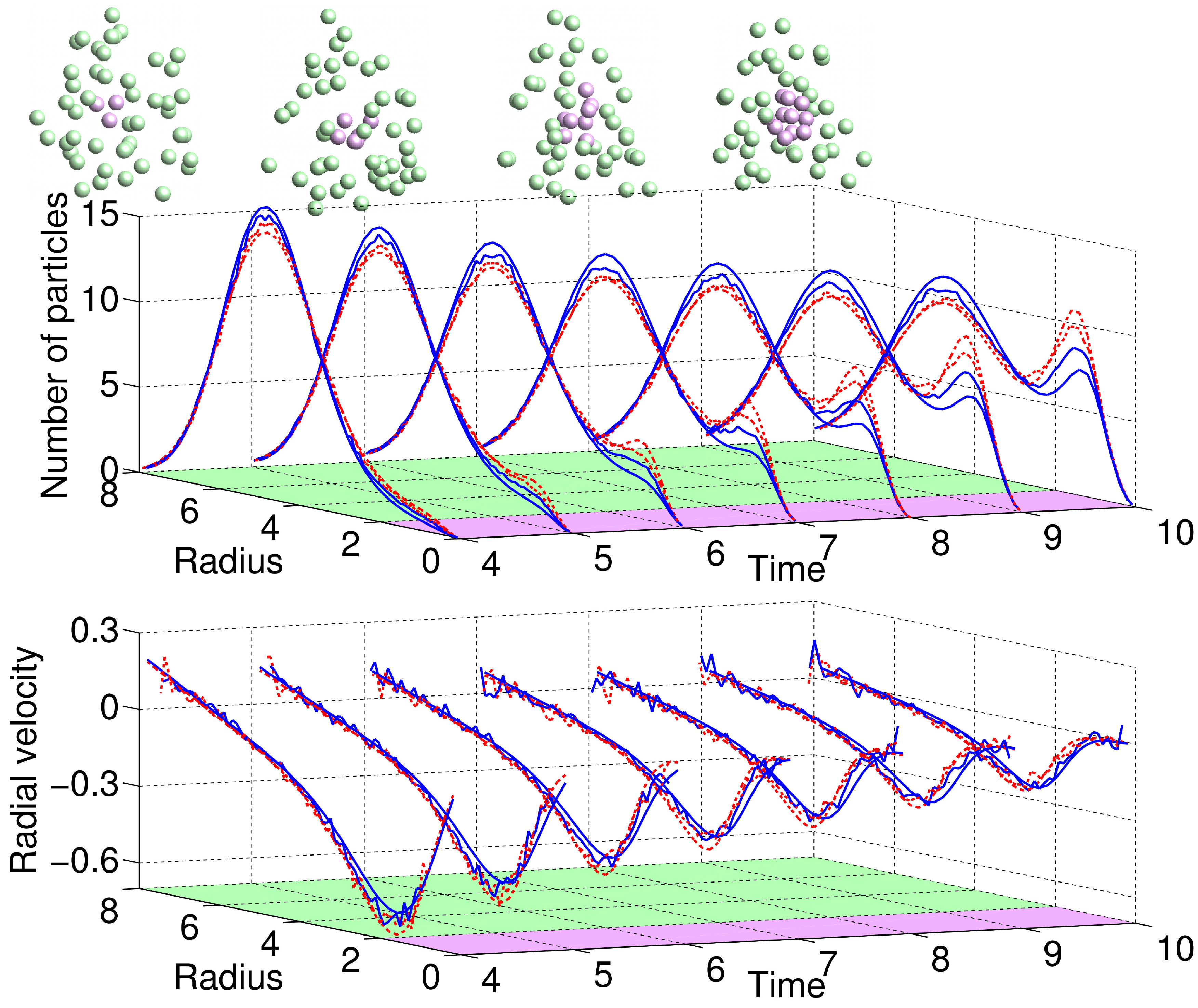}
\caption{ \label{Fig:50} Radial particle distribution and velocity
from DDFT \eqref{continuity} and \eqref{DDFT}
\emph{(smooth curves)} and stochastic equations \eqref{Newton}
\emph{(noisy curves)} with \emph{(blue, solid)} and without \emph{(red, dashes)}
HI. Also shown is one representative stochastic realization, at
times 4, 6, 8 and 10, 
particles coloured \emph{purple} for $|\bsr|<2$, \emph{green} otherwise. See
Supplemental
Movies 2 and 3.}
\end{figure}

\begin{figure}%[htb]
\includegraphics[width=0.45\textwidth]{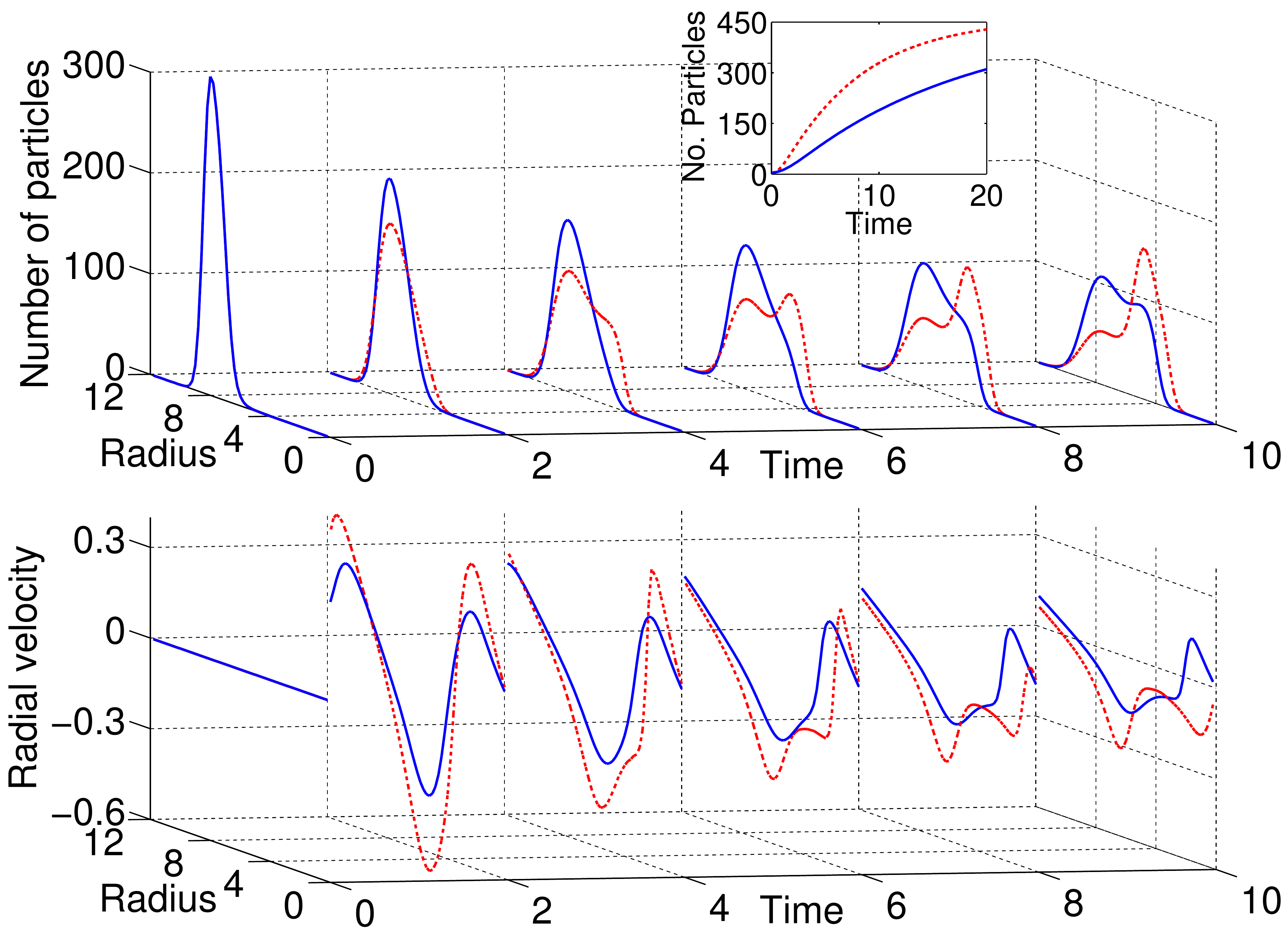}
\caption{ \label{Fig:500} Radial particle distribution and velocity
from DDFT \eqref{continuity} and \eqref{DDFT} with
\emph{(blue, solid)} and without \emph{(red, dashes)} HI.  
Inset: number of particles with $0<|\bsr|<6$.
See Supplemental Movie 4.
}
\end{figure}

Fig.~\ref{Fig:50} shows the evolution of the same 50 particles,
but we now switch
between potentials $V_1(r;6)$ and $V_1(r;0)$ only once, at time 0. We again have
very good agreement between our DDFT
and stochastic simulations. The small differences in the position distribution
near the origin are likely due to the choice of correlation
function, which is less accurate at higher densities. Here,
HI dramatically slow the build-up of
particles near the origin. Having verified our DDFT by comparison to
stochastic simulations, Fig.~\ref{Fig:500} 
shows the DDFT solution for 500 particles with the same potentials, for which
the stochastic equations are computationally very costly. 
The HI effects are even more dramatic, leading to
qualitatively different behaviour. This size-dependence shows that HI
must be carefully considered in any DDFT used to model macroscopic
numbers of particles.

From now on we consider two-body inter-particle potentials
and discuss
two limits of \eqref{DDFT}. Close to local equilibrium, we
expand $f^{(1)}$ and $f^{(2)}$ as Taylor series in
$\grad{\bsr_1}\bsv$ \ \cite{Kreuzer81}, obtaining a generalized
compressible, non-local Navier-Stokes-like integro-differential
equation:
\begin{align*}
     &\rho D_t \bsv
     =\eta \grad{\bsr_1}^2 \bsv +
    (\zeta + \tfrac{1}{3}\eta) \grad{\bsr_1}(\div{\bsr_1}\bsv) + \rho\,
\mathcal{G}([\rho],[\bsv]),
\end{align*}
where $\bsv=\bsv(\bsr_1,t)$, $\rho=\rho(\bsr_1,t)$ and
$\mathcal{G}([\rho],[\bsv])$ is the right hand side of \eqref{DDFT}.
The first three terms are standard but the viscosities $\eta$ and $\zeta$ are given by
integrals involving the two-body potential and the Taylor expansion
coefficients. Hence, the above equation
is not amenable to a straightforward numerical solution, as also is the case for
a simple
fluid~\cite{Kreuzer81}.
The new terms in $\mathcal{G}$ are a pressure-like term, depending on the
gradient of the
chemical potential, and HI terms, discussed above.

Most DDFTs are formulated in the high-friction regime, where the momenta of the colloidal particles
equilibrate on a much shorter timescale than their positions.   
In this regime, we have a non-dimensional parameter
$\epsilon=\sqrt{k_BT/m}\gamma^{-1}L^{-1}\ll 1$, where $L$ is a
`typical' length scale of the system. Denoting a Maxwellian momentum
distribution by
$M(\bsp_1)=\exp(-|\bsp_1|^2/(2mk_BT))/(2mk_BT\pi)^{3/2}$, we find
rigorously~\cite{GPK11} that $f^{(1)}(\bsr_1,\bsp_1,t)=M(\bsp_1)
(\rho(\bsr_1,t) + \epsilon \bsa(\bsr_1,t) \cdot \bsp_1 +
\mathcal{O}(\epsilon^2))$ for some function $\bsa$; in particular
$\int \dd \bsp_1 (\bsp_1 \otimes \bsp_1) f^{(1)}(\bsr_1,\bsp_1,t) =
\mathcal{O}(\epsilon^2)$. For ease of presentation, we
set $\bsZ_2=0$
(see~\cite{GPK11} for the generalization to $\bsZ_2 \neq 0$) and let
$\rho^{(2)}(\bsr_1,\bsr_2,t)=\rho(\bsr_1,t)\rho(\bsr_2,t)g(\bsr_1,\bsr_2)$.
Then $\rho$ satisfies a Smoluchowski
equation~\cite{GPK11}
with a novel diffusion tensor
\[
    \bsD(\bsr_1,[\rho]) = \frac{k_BT}{m\gamma} \Big[\bsone + \int \dd \bsr_2
g(\bsr_1,\bsr_2) \rho(\bsr_2,t)
\bsZ_1(\bsr_1,\bsr_2) \Big]^{-1},
\]
retained in the DDFT, cf.~\cite{RexLowen09}. 
Surprisingly, $\bsD$ is a non-local functional of $\rho$
and implicitly
time-dependent, even though the friction
tensor is time-independent.
Previous phenomenological attempts at including a density-dependent
diffusion coefficient in DDFTs do not correctly take into account
the form of the diffusion tensor~\cite{Rauscher10}.

Our new DDFT should accurately model a wide spectrum of
real-world problems and also
help elucidate the associated underlying phenomena. These
include systems in which HI or inertia are crucial, e.g. (i) wetting
phenomena~\cite{Peter2012,*Antonio2012,*Nikos2010};
(ii) transport and
coagulation of
nanoparticles in pulsatile and oscillatory
systems~\cite{WorthLongestKleinstreuer03};
and,
(iii) cloud formation and deposition of nanoparticles~
\cite{*WorthlongestXi07,*GavzeShapiro98, *PruppacherKlettWang98,
*SigurgeirssonStuart02,*FalkovichFouxonStepanov02}.
Furthermore, there are many promising extensions to the modelling
approach proposed here, e.g.\ to self-propelled particles, modelling
bacteria; multiple particle species; anisotropic particles; and the
inclusion of an external flow field, as would be required in
modelling blood and drug-laden nanoparticle movement in blood.
Similar approaches should also be highly relevant in
granular media, ion transport, and other multi-phase systems.

%\end{document}

\begin{acknowledgments}
We thank Petr Yatsyshin for stimulating discussions regarding
free-energy functionals. We are grateful to the European Research
Council via Advanced Grant No.\ 247031, the Rotary Clubs Darmstadt,
Darmstadt-Bergstra{\ss}e and Darmstadt-Kranichstein, the European
Framework 7 via Grant No.\ 214919 (Multiflow) and the Engineering and
Physical Sciences Research Council of the UK via grant No.\
EP/H034587/ for support of this research.
\end{acknowledgments}

% \bibliography{HIPRL}

\begin{thebibliography}{42}%
\makeatletter
\providecommand \@ifxundefined [1]{%
 \@ifx{#1\undefined}
}%
\providecommand \@ifnum [1]{%
 \ifnum #1\expandafter \@firstoftwo
 \else \expandafter \@secondoftwo
 \fi
}%
\providecommand \@ifx [1]{%
 \ifx #1\expandafter \@firstoftwo
 \else \expandafter \@secondoftwo
 \fi
}%
\providecommand \natexlab [1]{#1}%
\providecommand \enquote  [1]{``#1''}%
\providecommand \bibnamefont  [1]{#1}%
\providecommand \bibfnamefont [1]{#1}%
\providecommand \citenamefont [1]{#1}%
\providecommand \href@noop [0]{\@secondoftwo}%
\providecommand \href [0]{\begingroup \@sanitize@url \@href}%
\providecommand \@href[1]{\@@startlink{#1}\@@href}%
\providecommand \@@href[1]{\endgroup#1\@@endlink}%
\providecommand \@sanitize@url [0]{\catcode `\\12\catcode `\$12\catcode
  `\&12\catcode `\#12\catcode `\^12\catcode `\_12\catcode `\%12\relax}%
\providecommand \@@startlink[1]{}%
\providecommand \@@endlink[0]{}%
\providecommand \url  [0]{\begingroup\@sanitize@url \@url }%
\providecommand \@url [1]{\endgroup\@href {#1}{\urlprefix }}%
\providecommand \urlprefix  [0]{URL }%
\providecommand \Eprint [0]{\href }%
\providecommand \doibase [0]{http://dx.doi.org/}%
\providecommand \selectlanguage [0]{\@gobble}%
\providecommand \bibinfo  [0]{\@secondoftwo}%
\providecommand \bibfield  [0]{\@secondoftwo}%
\providecommand \translation [1]{[#1]}%
\providecommand \BibitemOpen [0]{}%
\providecommand \bibitemStop [0]{}%
\providecommand \bibitemNoStop [0]{.\EOS\space}%
\providecommand \EOS [0]{\spacefactor3000\relax}%
\providecommand \BibitemShut  [1]{\csname bibitem#1\endcsname}%
\let\auto@bib@innerbib\@empty
%</preamble>
\bibitem [{\citenamefont {Brown}(1828)}]{Brown28}%
  \BibitemOpen
  \bibfield  {author} {\bibinfo {author} {\bibfnamefont {R.}~\bibnamefont
  {Brown}},\ }\href@noop {} {\bibfield  {journal} {\bibinfo  {journal} {Phil.
  Mag.}\ }\textbf {\bibinfo {volume} {4}},\ \bibinfo {pages} {161} (\bibinfo
  {year} {1828})}\BibitemShut {NoStop}%
\bibitem [{\citenamefont {Einstein}(1905)}]{Einstein05}%
  \BibitemOpen
  \bibfield  {author} {\bibinfo {author} {\bibfnamefont {A.}~\bibnamefont
  {Einstein}},\ }\href@noop {} {\bibfield  {journal} {\bibinfo  {journal} {Ann.
  Phys. Lpz}\ }\textbf {\bibinfo {volume} {17}},\ \bibinfo {pages} {549}
  (\bibinfo {year} {1905})}\BibitemShut {NoStop}%
\bibitem [{\citenamefont {Langevin}(1908)}]{Langevin08}%
  \BibitemOpen
  \bibfield  {author} {\bibinfo {author} {\bibfnamefont {P.}~\bibnamefont
  {Langevin}},\ }\href@noop {} {\bibfield  {journal} {\bibinfo  {journal} {C.
  R. Acad. Sci. (Paris)}\ }\textbf {\bibinfo {volume} {146}},\ \bibinfo {pages}
  {530} (\bibinfo {year} {1908})}\BibitemShut {NoStop}%
\bibitem [{\citenamefont {Von~Smoluchowski}(1915)}]{Smoluchowski15}%
  \BibitemOpen
  \bibfield  {author} {\bibinfo {author} {\bibfnamefont {M.}~\bibnamefont
  {Von~Smoluchowski}},\ }\href@noop {} {\bibfield  {journal} {\bibinfo
  {journal} {Ann. Phys}\ }\textbf {\bibinfo {volume} {48}},\ \bibinfo {pages}
  {1103} (\bibinfo {year} {1915})}\BibitemShut {NoStop}%
\bibitem [{\citenamefont {Perrin}(1909)}]{Perrin09}%
  \BibitemOpen
  \bibfield  {author} {\bibinfo {author} {\bibfnamefont {J.}~\bibnamefont
  {Perrin}},\ }\href@noop {} {\bibfield  {journal} {\bibinfo  {journal} {Ann.
  Chim. Phys.}\ }\textbf {\bibinfo {volume} {18}},\ \bibinfo {pages} {1}
  (\bibinfo {year} {1909})}\BibitemShut {NoStop}%
\bibitem [{\citenamefont {Franosch}\ \emph {et~al.}(2011)\citenamefont
  {Franosch}, \citenamefont {Grimm}, \citenamefont {Belushkin}, \citenamefont
  {Mor}, \citenamefont {Foffi}, \citenamefont {Forr{\'o}},\ and\ \citenamefont
  {Jeney}}]{Franosch11}%
  \BibitemOpen
  \bibfield  {author} {\bibinfo {author} {\bibfnamefont {T.}~\bibnamefont
  {Franosch}}, \bibinfo {author} {\bibfnamefont {M.}~\bibnamefont {Grimm}},
  \bibinfo {author} {\bibfnamefont {M.}~\bibnamefont {Belushkin}}, \bibinfo
  {author} {\bibfnamefont {F.}~\bibnamefont {Mor}}, \bibinfo {author}
  {\bibfnamefont {G.}~\bibnamefont {Foffi}}, \bibinfo {author} {\bibfnamefont
  {L.}~\bibnamefont {Forr{\'o}}}, \ and\ \bibinfo {author} {\bibfnamefont
  {S.}~\bibnamefont {Jeney}},\ }\href@noop {} {\bibfield  {journal} {\bibinfo
  {journal} {Nature}\ }\textbf {\bibinfo {volume} {478}},\ \bibinfo {pages}
  {85} (\bibinfo {year} {2011})}\BibitemShut {NoStop}%
\bibitem [{\citenamefont {Huang}\ \emph {et~al.}(2011)\citenamefont {Huang},
  \citenamefont {Chavez}, \citenamefont {Taute}, \citenamefont {Luki{\'c}},
  \citenamefont {Jeney}, \citenamefont {Raizen},\ and\ \citenamefont
  {Florin}}]{Huang11}%
  \BibitemOpen
  \bibfield  {author} {\bibinfo {author} {\bibfnamefont {R.}~\bibnamefont
  {Huang}}, \bibinfo {author} {\bibfnamefont {I.}~\bibnamefont {Chavez}},
  \bibinfo {author} {\bibfnamefont {K.}~\bibnamefont {Taute}}, \bibinfo
  {author} {\bibfnamefont {B.}~\bibnamefont {Luki{\'c}}}, \bibinfo {author}
  {\bibfnamefont {S.}~\bibnamefont {Jeney}}, \bibinfo {author} {\bibfnamefont
  {M.~G.}\ \bibnamefont {Raizen}}, \ and\ \bibinfo {author} {\bibfnamefont
  {E.~L.}\ \bibnamefont {Florin}},\ }\href@noop {} {\bibfield  {journal}
  {\bibinfo  {journal} {Nature Physics}\ }\textbf {\bibinfo {volume} {7}},\
  \bibinfo {pages} {576} (\bibinfo {year} {2011})}\BibitemShut {NoStop}%
\bibitem [{\citenamefont {Matijevi{\'c}}(2008)}]{Matijevic08}%
  \BibitemOpen
  \bibinfo {editor} {\bibfnamefont {E.}~\bibnamefont {Matijevi{\'c}}},\ ed.,\
  \href@noop {} {\emph {\bibinfo {title} {{Medical Applications of
  Colloids}}}}\ (\bibinfo  {publisher} {Springer, New York},\ \bibinfo {year}
  {2008})\BibitemShut {NoStop}%
\bibitem [{\citenamefont {Evans}\ and\ \citenamefont
  {Wennerstr{\"o}m}(1999)}]{Evans99}%
  \BibitemOpen
  \bibfield  {author} {\bibinfo {author} {\bibfnamefont {D.~F.}\ \bibnamefont
  {Evans}}\ and\ \bibinfo {author} {\bibfnamefont {H.}~\bibnamefont
  {Wennerstr{\"o}m}},\ }\href@noop {} {\emph {\bibinfo {title} {{The Colloidal
  Domain: Where Physics, Chemistry, and Biology Meet}}}}\ (\bibinfo
  {publisher} {Wiley, New York},\ \bibinfo {year} {1999})\BibitemShut {NoStop}%
\bibitem [{\citenamefont {Stone}\ and\ \citenamefont {Kim}(2001)}]{Stone2001}%
  \BibitemOpen
  \bibfield  {author} {\bibinfo {author} {\bibfnamefont {H.}~\bibnamefont
  {Stone}}\ and\ \bibinfo {author} {\bibfnamefont {S.}~\bibnamefont {Kim}},\
  }\href@noop {} {\bibfield  {journal} {\bibinfo  {journal} {AIChE J.}\
  }\textbf {\bibinfo {volume} {47}},\ \bibinfo {pages} {1250} (\bibinfo {year}
  {2001})}\BibitemShut {NoStop}%
\bibitem [{\citenamefont {Caruso}(2004)}]{Caruso2004}%
  \BibitemOpen
  \bibfield  {author} {\bibinfo {author} {\bibfnamefont {F.}~\bibnamefont
  {Caruso}},\ }\href@noop {} {\emph {\bibinfo {title} {{Colloids and Colloid
  Assemblies: Synthesis, Modification, Organization and Utilization of Colloid
  Particles}}}}\ (\bibinfo  {publisher} {Wiley, Weinheim},\ \bibinfo {year}
  {2004})\BibitemShut {NoStop}%
\bibitem [{\citenamefont {Kreuzer}(1981)}]{Kreuzer81}%
  \BibitemOpen
  \bibfield  {author} {\bibinfo {author} {\bibfnamefont {H.~J.}\ \bibnamefont
  {Kreuzer}},\ }\href@noop {} {\emph {\bibinfo {title} {{Nonequilibrium
  Thermodynamics and Its Statistical Foundations}}}}\ (\bibinfo  {publisher}
  {Oxford University Press, New York},\ \bibinfo {year} {1981})\BibitemShut
  {NoStop}%
\bibitem [{\citenamefont {Lutsko}(2010)}]{Lutsko10}%
  \BibitemOpen
  \bibfield  {author} {\bibinfo {author} {\bibfnamefont {J.~F.}\ \bibnamefont
  {Lutsko}},\ }\href@noop {} {\bibfield  {journal} {\bibinfo  {journal} {Adv.
  Chem. Phys.}\ }\textbf {\bibinfo {volume} {144}},\ \bibinfo {pages} {1}
  (\bibinfo {year} {2010})}\BibitemShut {NoStop}%
\bibitem [{\citenamefont {Rex}\ and\ \citenamefont
  {L\"owen}(2009)}]{RexLowen09}%
  \BibitemOpen
  \bibfield  {author} {\bibinfo {author} {\bibfnamefont {M.}~\bibnamefont
  {Rex}}\ and\ \bibinfo {author} {\bibfnamefont {H.}~\bibnamefont {L\"owen}},\
  }\href@noop {} {\bibfield  {journal} {\bibinfo  {journal} {Eur. Phys. J. E}\
  }\textbf {\bibinfo {volume} {28}},\ \bibinfo {pages} {139} (\bibinfo {year}
  {2009})}\BibitemShut {NoStop}%
\bibitem [{\citenamefont {Archer}(2009)}]{Archer09}%
  \BibitemOpen
  \bibfield  {author} {\bibinfo {author} {\bibfnamefont {A.~J.}\ \bibnamefont
  {Archer}},\ }\href@noop {} {\bibfield  {journal} {\bibinfo  {journal} {J.
  Chem. Phys.}\ }\textbf {\bibinfo {volume} {130}},\ \bibinfo {pages} {014509}
  (\bibinfo {year} {2009})}\BibitemShut {NoStop}%
\bibitem [{\citenamefont {Marconi}\ and\ \citenamefont
  {Tarazona}(1999)}]{MarconiTarazona99}%
  \BibitemOpen
  \bibfield  {author} {\bibinfo {author} {\bibfnamefont {U.~M.~B.}\
  \bibnamefont {Marconi}}\ and\ \bibinfo {author} {\bibfnamefont
  {P.}~\bibnamefont {Tarazona}},\ }\href@noop {} {\bibfield  {journal}
  {\bibinfo  {journal} {J. Chem. Phys.}\ }\textbf {\bibinfo {volume} {110}},\
  \bibinfo {pages} {8032} (\bibinfo {year} {1999})}\BibitemShut {NoStop}%
\bibitem [{\citenamefont {Goddard}\ \emph {et~al.}(2012)\citenamefont
  {Goddard}, \citenamefont {Pavliotis},\ and\ \citenamefont
  {Kalliadasis}}]{GPK11}%
  \BibitemOpen
  \bibfield  {author} {\bibinfo {author} {\bibfnamefont {B.~D.}\ \bibnamefont
  {Goddard}}, \bibinfo {author} {\bibfnamefont {G.~A.}\ \bibnamefont
  {Pavliotis}}, \ and\ \bibinfo {author} {\bibfnamefont {S.}~\bibnamefont
  {Kalliadasis}},\ }\href@noop {} {\bibfield  {journal} {\bibinfo  {journal}
  {Multiscale Model. Sim.}\ }\textbf {\bibinfo {volume} {10}},\ \bibinfo
  {pages} {633} (\bibinfo {year} {2012})}\BibitemShut {NoStop}%
\bibitem [{\citenamefont {Dhont}(1996)}]{Dhont96}%
  \BibitemOpen
  \bibfield  {author} {\bibinfo {author} {\bibfnamefont {J.~K.~G.}\
  \bibnamefont {Dhont}},\ }\href@noop {} {\emph {\bibinfo {title} {{An
  Introduction to Dynamics of Colloids}}}}\ (\bibinfo  {publisher} {Elsevier},\
  \bibinfo {year} {1996})\BibitemShut {NoStop}%
\bibitem [{\citenamefont {Padding}\ and\ \citenamefont
  {Louis}(2006)}]{PaddingLouis06}%
  \BibitemOpen
  \bibfield  {author} {\bibinfo {author} {\bibfnamefont {J.~T.}\ \bibnamefont
  {Padding}}\ and\ \bibinfo {author} {\bibfnamefont {A.~A.}\ \bibnamefont
  {Louis}},\ }\href@noop {} {\bibfield  {journal} {\bibinfo  {journal} {Phys.
  Rev. E}\ }\textbf {\bibinfo {volume} {74}},\ \bibinfo {pages} {031402}
  (\bibinfo {year} {2006})}\BibitemShut {NoStop}%
\bibitem [{\citenamefont {Ermak}\ and\ \citenamefont
  {McCammon}(1978)}]{ErmakMcCammon78}%
  \BibitemOpen
  \bibfield  {author} {\bibinfo {author} {\bibfnamefont {D.~L.}\ \bibnamefont
  {Ermak}}\ and\ \bibinfo {author} {\bibfnamefont {J.~A.}\ \bibnamefont
  {McCammon}},\ }\href@noop {} {\bibfield  {journal} {\bibinfo  {journal} {J.
  Chem. Phys.}\ }\textbf {\bibinfo {volume} {69}},\ \bibinfo {pages} {1351}
  (\bibinfo {year} {1978})}\BibitemShut {NoStop}%
\bibitem [{\citenamefont {Chan}\ and\ \citenamefont
  {Finken}(2005)}]{ChanFinken05}%
  \BibitemOpen
  \bibfield  {author} {\bibinfo {author} {\bibfnamefont {G.~K.-L.}\
  \bibnamefont {Chan}}\ and\ \bibinfo {author} {\bibfnamefont {R.}~\bibnamefont
  {Finken}},\ }\href@noop {} {\bibfield  {journal} {\bibinfo  {journal} {Phys.
  Rev. Lett.}\ }\textbf {\bibinfo {volume} {94}},\ \bibinfo {pages} {183001}
  (\bibinfo {year} {2005})}\BibitemShut {NoStop}%
\bibitem [{\citenamefont {Marques}\ \emph {et~al.}(2006)\citenamefont
  {Marques}, \citenamefont {Ullrich}, \citenamefont {Nogueira}, \citenamefont
  {Rubio}, \citenamefont {Burke},\ and\ \citenamefont {Gross}}]{Marques06}%
  \BibitemOpen
  \bibinfo {editor} {\bibfnamefont {M.~A.~L.}\ \bibnamefont {Marques}},
  \bibinfo {editor} {\bibfnamefont {C.~A.}\ \bibnamefont {Ullrich}}, \bibinfo
  {editor} {\bibfnamefont {F.}~\bibnamefont {Nogueira}}, \bibinfo {editor}
  {\bibfnamefont {A.}~\bibnamefont {Rubio}}, \bibinfo {editor} {\bibfnamefont
  {K.}~\bibnamefont {Burke}}, \ and\ \bibinfo {editor} {\bibfnamefont
  {E.~K.~U.}\ \bibnamefont {Gross}},\ eds.,\ \href@noop {} {\emph {\bibinfo
  {title} {{Time-Dependent Density Functional Theory}}}},\ \bibinfo {series}
  {Lect. Notes Phys.}, Vol.\ \bibinfo {volume} {706}\ (\bibinfo  {publisher}
  {Springer, Berlin},\ \bibinfo {year} {2006})\BibitemShut {NoStop}%
\bibitem [{\citenamefont {R{\'e}sibois}\ and\ \citenamefont
  {De~Leener}(1977)}]{ResiboisDeLeener77}%
  \BibitemOpen
  \bibfield  {author} {\bibinfo {author} {\bibfnamefont {P.}~\bibnamefont
  {R{\'e}sibois}}\ and\ \bibinfo {author} {\bibfnamefont {M.}~\bibnamefont
  {De~Leener}},\ }\href@noop {} {\emph {\bibinfo {title} {Classical Kinetic
  Theory of Fluids}}}\ (\bibinfo  {publisher} {Wiley},\ \bibinfo {year}
  {1977})\BibitemShut {NoStop}%
\bibitem [{\citenamefont {Knudsen}\ \emph {et~al.}(2008)\citenamefont
  {Knudsen}, \citenamefont {Werth},\ and\ \citenamefont
  {Wolf}}]{KnudsenWerthWolf08}%
  \BibitemOpen
  \bibfield  {author} {\bibinfo {author} {\bibfnamefont {H.~A.}\ \bibnamefont
  {Knudsen}}, \bibinfo {author} {\bibfnamefont {J.~H.}\ \bibnamefont {Werth}},
  \ and\ \bibinfo {author} {\bibfnamefont {D.~E.}\ \bibnamefont {Wolf}},\
  }\href@noop {} {\bibfield  {journal} {\bibinfo  {journal} {Eur. Phys. J. E}\
  }\textbf {\bibinfo {volume} {27}},\ \bibinfo {pages} {161} (\bibinfo {year}
  {2008})}\BibitemShut {NoStop}%
\bibitem [{\citenamefont {Evans}(1979)}]{Evans79}%
  \BibitemOpen
  \bibfield  {author} {\bibinfo {author} {\bibfnamefont {R.}~\bibnamefont
  {Evans}},\ }\href@noop {} {\bibfield  {journal} {\bibinfo  {journal} {Adv.
  Phys.}\ }\textbf {\bibinfo {volume} {28}},\ \bibinfo {pages} {143} (\bibinfo
  {year} {1979})}\BibitemShut {NoStop}%
\bibitem [{\citenamefont {Rosenfeld}(1989)}]{Rosenfeld89}%
  \BibitemOpen
  \bibfield  {author} {\bibinfo {author} {\bibfnamefont {Y.}~\bibnamefont
  {Rosenfeld}},\ }\href@noop {} {\bibfield  {journal} {\bibinfo  {journal}
  {Phys. Rev. Lett.}\ }\textbf {\bibinfo {volume} {63}},\ \bibinfo {pages}
  {980} (\bibinfo {year} {1989})}\BibitemShut {NoStop}%
\bibitem [{\citenamefont {Hughes}\ and\ \citenamefont
  {Burghardt}(2012)}]{HughesBurghardt12}%
  \BibitemOpen
  \bibfield  {author} {\bibinfo {author} {\bibfnamefont {K.~H.}\ \bibnamefont
  {Hughes}}\ and\ \bibinfo {author} {\bibfnamefont {I.}~\bibnamefont
  {Burghardt}},\ }\href@noop {} {\bibfield  {journal} {\bibinfo  {journal} {J.
  Chem. Phys.}\ }\textbf {\bibinfo {volume} {136}},\ \bibinfo {pages} {214109}
  (\bibinfo {year} {2012})}\BibitemShut {NoStop}%
\bibitem [{\citenamefont {Rotne}\ and\ \citenamefont
  {Prager}(1969)}]{RotnePrager69}%
  \BibitemOpen
  \bibfield  {author} {\bibinfo {author} {\bibfnamefont {J.}~\bibnamefont
  {Rotne}}\ and\ \bibinfo {author} {\bibfnamefont {S.}~\bibnamefont {Prager}},\
  }\href@noop {} {\bibfield  {journal} {\bibinfo  {journal} {J. Chem. Phys.}\
  }\textbf {\bibinfo {volume} {50}},\ \bibinfo {pages} {4831} (\bibinfo {year}
  {1969})}\BibitemShut {NoStop}%
\bibitem [{\citenamefont {Jeffrey}\ and\ \citenamefont
  {Onishi}(1984)}]{JeffreyOnishi84}%
  \BibitemOpen
  \bibfield  {author} {\bibinfo {author} {\bibfnamefont {D.~J.}\ \bibnamefont
  {Jeffrey}}\ and\ \bibinfo {author} {\bibfnamefont {Y.}~\bibnamefont
  {Onishi}},\ }\href@noop {} {\bibfield  {journal} {\bibinfo  {journal} {J.
  Fluid Mech.}\ }\textbf {\bibinfo {volume} {139}},\ \bibinfo {pages} {261}
  (\bibinfo {year} {1984})}\BibitemShut {NoStop}%
\bibitem [{\citenamefont {Boyd}(2001)}]{Boyd01}%
  \BibitemOpen
  \bibfield  {author} {\bibinfo {author} {\bibfnamefont {J.~P.}\ \bibnamefont
  {Boyd}},\ }\href@noop {} {\emph {\bibinfo {title} {{Chebyshev and Fourier
  Spectral Methods}}}}\ (\bibinfo  {publisher} {Dover, UK},\ \bibinfo {year}
  {2001})\BibitemShut {NoStop}%
\bibitem [{\citenamefont {Hairer}\ and\ \citenamefont
  {Wanner}(2006)}]{HairerWanner96}%
  \BibitemOpen
  \bibfield  {author} {\bibinfo {author} {\bibfnamefont {E.}~\bibnamefont
  {Hairer}}\ and\ \bibinfo {author} {\bibfnamefont {G.}~\bibnamefont
  {Wanner}},\ }\href@noop {} {\emph {\bibinfo {title} {{Solving Ordinary
  Differential Equations {II}. Stiff and Differential-Algebraic Problems}}}},\
  \bibinfo {series} {Springer Series in Computational Mathematics},
  Vol.~\bibinfo {volume} {14}\ (\bibinfo  {publisher} {Springer-Verlag,
  Berlin},\ \bibinfo {year} {2006})\BibitemShut {NoStop}%
\bibitem [{\citenamefont {Kloeden}\ and\ \citenamefont
  {Platen}(1992)}]{KloedenPlaten92}%
  \BibitemOpen
  \bibfield  {author} {\bibinfo {author} {\bibfnamefont {P.~E.}\ \bibnamefont
  {Kloeden}}\ and\ \bibinfo {author} {\bibfnamefont {E.}~\bibnamefont
  {Platen}},\ }\href@noop {} {\emph {\bibinfo {title} {{Numerical Solution of
  Stochastic Differential Equations}}}}\ (\bibinfo  {publisher}
  {Springer-Verlag, Berlin},\ \bibinfo {year} {1992})\BibitemShut {NoStop}%
\bibitem [{\citenamefont {Rauscher}(2010)}]{Rauscher10}%
  \BibitemOpen
  \bibfield  {author} {\bibinfo {author} {\bibfnamefont {M.}~\bibnamefont
  {Rauscher}},\ }\href@noop {} {\bibfield  {journal} {\bibinfo  {journal} {J.
  Phys. Condens. Matter}\ }\textbf {\bibinfo {volume} {22}},\ \bibinfo {pages}
  {364109} (\bibinfo {year} {2010})}\BibitemShut {NoStop}%
\bibitem [{\citenamefont {Yatsyshin}\ \emph {et~al.}(2012)\citenamefont
  {Yatsyshin}, \citenamefont {Savva},\ and\ \citenamefont
  {Kalliadasis}}]{Peter2012}%
  \BibitemOpen
  \bibfield  {author} {\bibinfo {author} {\bibfnamefont {P.}~\bibnamefont
  {Yatsyshin}}, \bibinfo {author} {\bibfnamefont {N.}~\bibnamefont {Savva}}, \
  and\ \bibinfo {author} {\bibfnamefont {S.}~\bibnamefont {Kalliadasis}},\
  }\href@noop {} {\bibfield  {journal} {\bibinfo  {journal} {J. Chem. Phys.}\
  }\textbf {\bibinfo {volume} {136}},\ \bibinfo {pages} {124113} (\bibinfo
  {year} {2012})}\BibitemShut {NoStop}%
\bibitem [{\citenamefont {Pereira}\ and\ \citenamefont
  {Kalliadasis}(2012)}]{Antonio2012}%
  \BibitemOpen
  \bibfield  {author} {\bibinfo {author} {\bibfnamefont {A.}~\bibnamefont
  {Pereira}}\ and\ \bibinfo {author} {\bibfnamefont {S.}~\bibnamefont
  {Kalliadasis}},\ }\href@noop {} {\bibfield  {journal} {\bibinfo  {journal}
  {J. Fluid Mech.}\ }\textbf {\bibinfo {volume} {692}},\ \bibinfo {pages} {53}
  (\bibinfo {year} {2012})}\BibitemShut {NoStop}%
\bibitem [{\citenamefont {Savva}\ \emph {et~al.}(2010)\citenamefont {Savva},
  \citenamefont {Kalliadasis},\ and\ \citenamefont {Pavliotis}}]{Nikos2010}%
  \BibitemOpen
  \bibfield  {author} {\bibinfo {author} {\bibfnamefont {N.}~\bibnamefont
  {Savva}}, \bibinfo {author} {\bibfnamefont {S.}~\bibnamefont {Kalliadasis}},
  \ and\ \bibinfo {author} {\bibfnamefont {G.}~\bibnamefont {Pavliotis}},\
  }\href@noop {} {\bibfield  {journal} {\bibinfo  {journal} {Phys. Rev. Lett.}\
  }\textbf {\bibinfo {volume} {104}},\ \bibinfo {pages} {084501} (\bibinfo
  {year} {2010})}\BibitemShut {NoStop}%
\bibitem [{\citenamefont {Worth~Longest}\ and\ \citenamefont
  {Kleinstreuer}(2003)}]{WorthLongestKleinstreuer03}%
  \BibitemOpen
  \bibfield  {author} {\bibinfo {author} {\bibfnamefont {P.}~\bibnamefont
  {Worth~Longest}}\ and\ \bibinfo {author} {\bibfnamefont {C.}~\bibnamefont
  {Kleinstreuer}},\ }\href@noop {} {\bibfield  {journal} {\bibinfo  {journal}
  {J. Biomech.}\ }\textbf {\bibinfo {volume} {36}},\ \bibinfo {pages} {421 }
  (\bibinfo {year} {2003})}\BibitemShut {NoStop}%
\bibitem [{\citenamefont {Worth~Longest}\ and\ \citenamefont
  {Xi}(2007)}]{WorthlongestXi07}%
  \BibitemOpen
  \bibfield  {author} {\bibinfo {author} {\bibfnamefont {P.}~\bibnamefont
  {Worth~Longest}}\ and\ \bibinfo {author} {\bibfnamefont {J.}~\bibnamefont
  {Xi}},\ }\href@noop {} {\bibfield  {journal} {\bibinfo  {journal} {J. Aerosol
  Sci.}\ }\textbf {\bibinfo {volume} {38}},\ \bibinfo {pages} {111} (\bibinfo
  {year} {2007})}\BibitemShut {NoStop}%
\bibitem [{\citenamefont {Gavze}\ and\ \citenamefont
  {Shapiro}(1998)}]{GavzeShapiro98}%
  \BibitemOpen
  \bibfield  {author} {\bibinfo {author} {\bibfnamefont {E.}~\bibnamefont
  {Gavze}}\ and\ \bibinfo {author} {\bibfnamefont {M.}~\bibnamefont
  {Shapiro}},\ }\href@noop {} {\bibfield  {journal} {\bibinfo  {journal} {J.
  Fluid Mech.}\ }\textbf {\bibinfo {volume} {371}},\ \bibinfo {pages} {59}
  (\bibinfo {year} {1998})}\BibitemShut {NoStop}%
\bibitem [{\citenamefont {Pruppacher}\ \emph {et~al.}(1998)\citenamefont
  {Pruppacher}, \citenamefont {Klett},\ and\ \citenamefont
  {Wang}}]{PruppacherKlettWang98}%
  \BibitemOpen
  \bibfield  {author} {\bibinfo {author} {\bibfnamefont {H.~R.}\ \bibnamefont
  {Pruppacher}}, \bibinfo {author} {\bibfnamefont {J.~D.}\ \bibnamefont
  {Klett}}, \ and\ \bibinfo {author} {\bibfnamefont {P.~K.}\ \bibnamefont
  {Wang}},\ }\href@noop {} {\emph {\bibinfo {title} {Microphysics of clouds and
  precipitation}}}\ (\bibinfo  {publisher} {Taylor \& Francis},\ \bibinfo
  {year} {1998})\BibitemShut {NoStop}%
\bibitem [{\citenamefont {Sigurgeirsson}\ and\ \citenamefont
  {Stuart}(2002)}]{SigurgeirssonStuart02}%
  \BibitemOpen
  \bibfield  {author} {\bibinfo {author} {\bibfnamefont {H.}~\bibnamefont
  {Sigurgeirsson}}\ and\ \bibinfo {author} {\bibfnamefont {A.~M.}\ \bibnamefont
  {Stuart}},\ }\href@noop {} {\bibfield  {journal} {\bibinfo  {journal} {Phys.
  Fluids}\ }\textbf {\bibinfo {volume} {14}},\ \bibinfo {pages} {4352}
  (\bibinfo {year} {2002})}\BibitemShut {NoStop}%
\bibitem [{\citenamefont {Falkovich}\ \emph {et~al.}(2002)\citenamefont
  {Falkovich}, \citenamefont {Fouxon},\ and\ \citenamefont
  {Stepanov}}]{FalkovichFouxonStepanov02}%
  \BibitemOpen
  \bibfield  {author} {\bibinfo {author} {\bibfnamefont {G.}~\bibnamefont
  {Falkovich}}, \bibinfo {author} {\bibfnamefont {A.}~\bibnamefont {Fouxon}}, \
  and\ \bibinfo {author} {\bibfnamefont {M.}~\bibnamefont {Stepanov}},\
  }\href@noop {} {\bibfield  {journal} {\bibinfo  {journal} {Nature}\ }\textbf
  {\bibinfo {volume} {419}},\ \bibinfo {pages} {151} (\bibinfo {year}
  {2002})}\BibitemShut {NoStop}%
\end{thebibliography}

%

\end{document}